\documentclass[sigconf]{acmart}

\setcopyright{acmcopyright}
\copyrightyear{2018}
\acmYear{2018}
\acmDOI{XXXXXXX.XXXXXXX}

\acmConference[ICSE '24]{Make sure to enter the correct
  conference title from your rights confirmation emai}{June 03--05,
  2018}{Woodstock, NY}

  \acmPrice{15.00}
\acmISBN{978-1-4503-XXXX-X/18/06}

\usepackage[english]{babel}

\usepackage{amsmath}
\usepackage{graphicx}
\usepackage{hyperref}

\usepackage{todonotes}
\usepackage{csquotes}

\begin{document}

\title{Challenges, Strengths, and Strategies of Software Engineers with ADHD: A Case Study}

\author{Grischa Liebel}
\email{grischal@ru.is}
\orcid{0000-0002-3884-815X}
\affiliation{\institution{School of Technology, Reykjavik University}
  \city{Reykjavik}
  \country{Iceland}
}

\author{Noah Langlois}
\email{noahlanglois@hotmail.com}
\affiliation{\institution{ISAE-ENSMA}
  \streetaddress{}
  \city{Poitiers}
  \country{France}
  \postcode{}
}

\author{Kiev Gama}
\email{kiev@cin.ufpe.br}
\affiliation{\institution{Federal University of Pernambuco (UFPE)}
  \city{Recife}
  \country{Brazil}
}

\renewcommand{\shortauthors}{Liebel et al.}

\begin{abstract}
Neurodiversity describes brain function variation in individuals, including Attention deficit hyperactivity disorder (ADHD) and Autism spectrum disorder.
Neurodivergent individuals both experience challenges and exhibit strengths in the workplace.
As an important disorder included under the neurodiversity term, an estimated 5.0\% to 7.1\% of the world population have ADHD. 
However, existing studies involving ADHD in the workplace are of general nature and do not focus on software engineering (SE) activities.
To address this gap, we performed an exploratory qualitative case study on the experiences of people with ADHD working in SE.
We find that people with ADHD struggle with several important SE-related activities, e.g., task organisation and estimation, attention to work, relation to others.
Furthermore, they experience issues with physical and mental health.
In terms of strengths, they exhibit, e.g., increased creative skills, perform well when solving puzzles, and have the capability to think ahead. 
Our findings align well with existing clinical ADHD research, and have important implications to SE practice.
\\
\textit{Lay Abstract --} Neurodiversity describes brain function variation in individuals, such as Attention deficit hyperactivity disorder (ADHD) and Autism spectrum disorder.
People included under this term often experience problems in their work, e.g., due to differences in communication or behaviour, but also exhibit strengths compared to people without these disorders.
To better include them, it is essential that we understand how these challenges and strengths manifest in different professions.
There is currently only initial research on how neurodiversity affects professionals in software engineering (SE), an environment characterised by a rapid pace, frequent change, and intense collaborative work.
Therefore, we studied the strengths, challenges, and strategies of SE professionals with ADHD, a disorder affecting approximately 5.0\% to 7.1\% of the world population.
We find that these professionals perceive many common SE activities as challenging, e.g., estimating how long tasks take, or maintaining focus.
Interestingly, they also exhibit a number of strengths that are highly relevant to the SE industry, such as increased creativity and systems thinking.
Based on our findings, we provide several recommendations on how SE companies can better support employees with ADHD.
\end{abstract}



\maketitle

\section{Introduction}
Neurodiversity describes variation in brain function among individuals, including conditions such as Attention deficit hyperactivity disorder (ADHD) and Autism spectrum disorder (ASD). 
Neurodivergent individuals face substantial barriers in society, e.g., due to difficulties or differences in communication, reading or writing difficulties, or reduced attention span \cite{ucu_workplace}. 
However, the concept of neurodiversity emphasises the value and strengths of individuals with neurological differences rather than focusing on their weaknesses~\cite{doyle2020neurodiversity}.

While computer-related jobs are often considered fitting for neurodivergent individuals or matching their interest profiles~\cite{austin2017neurodiversity}, there is only initial understanding of how they perform in software engineering (SE) \cite{morris15,gama2023understanding}. These preliminary studies provide a broader perspective on neurodiversity, without investigating more deeply any of those disorders.
To improve inclusion in SE, the particular challenges and strengths of each group need to be better understood.
For instance, in terms of creativity, individuals who report higher rates of ADHD symptoms outperform those with lower self-reported ADHD symptoms in divergent thinking tasks~\cite{hoogman2020creativity}.

As an important disorder included under the neurodiversity term, an estimated 5.0\% to 7.1\% of the world population \cite{willcutt2012prevalence,polanczyk2007worldwide} have ADHD. In the 2022 Stack Overflow Developer Survey~\cite{stackoverflow}, the top rank on the neurodiversity question consisted of 10.57\% of respondents (70K developers) indicating they have a concentration and/or memory disorder (e.g., ADHD).
Although limited compared to studies with children, research on ADHD in adults has grown in recent decades, with strong evidence suggesting that ADHD manifests differently in adults and children~\cite{barkley2010adhd}. This leads to struggles during college, professional activities and personal relationships~\cite{fleming2012developmental,barkley2010adhd}. Adults with ADHD often struggle to keep a steady job and tend to switch jobs more often, either because of layoffs due to low-rated work performance or due to impulsivity in quitting or switching jobs~\cite{fleming2012developmental}. Studies involving ADHD in the workplace, e.g., \cite{sarkis2014addressing, doyle2020neurodiversity}, are typically of general nature and do not focus on SE activities. 
To address this gap, we performed an exploratory qualitative case study on the experiences of people with ADHD working in SE. Our goal was to understand their challenges, strengths, and the strategies they use to address their challenges. Specifically, we answer the following research questions:
\begin{enumerate}
    \item[RQ1:] What challenges and strengths do people with ADHD experience in SE-related activities? 
    \item[RQ2:] What strategies do people with ADHD apply to cope with their challenges? 
\end{enumerate}
We extracted 15 challenges, 8 strengths, and 12 strategies from 19 interviews and discussed their implications with 4 experienced managers working in SE. We found that our interviewees struggle with task organisation and estimation, attention to work, relation to others, and with physical and mental health.
In terms of strengths, they, e.g., exhibit increased creative skills, perform well when solving puzzles or exploring novel topics, have the tendency to have broad interests and the capability to think ahead. 
These findings are aligned with clinical literature on ADHD and we map them to the software engineering practice. The interview instruments as well as six anonymised interview transcripts are available in the dataset\footnote{\url{https://dx.doi.org/10.5281/zenodo.8414120}} accompanying this publication.

\section{Background \& Related Work}

\subsection{Executive Functions and Self-Regulation}
ADHD is a long-lasting condition that affects various aspects of life from childhood to adulthood.
Recent understandings see it as a disorder of self-regulation and its underlying executive functioning~\cite{barkley2006attention}. Executive function (EF) is an umbrella term used for a range of hypothesised cognitive processes, including planning, working memory, attention, inhibition~\cite{Goldstein2014}. These cognitive processes have been classified as ``cool'' EFs, while ``hot'' EFs are related to emotional and affective aspects of cognition that also involve rewards and motivations~\cite{zelazo2002executive}. For instance, this leads to the typical characteristic of delay aversion~\cite{sonuga1992hyperactivity} found in many individuals with ADHD, i.e., a preference for immediate over delayed rewards. Zelazo and M{\"u}ller~\cite{zelazo2002executive} state that ADHD literature often conflates hot and cool EF or considers EF and cool EF to be synonyms. Popular symptoms of ADHD are often limited to a perspective around hyperactivity and attention, mostly because of the acronym and the DSM-5~\cite{american2013diagnostic}\footnote{The Diagnostic and Statistical Manual (DSM) of Mental Disorders, by the American Psychiatric Association (APA), serves as the principal authority for psychiatric diagnoses. Although being used worldwide, it has attracted criticism about its role in processes of medicalization of mental illness~\cite{pickersgill2014debating}}. Such simplification on the perspective of EF limited to cognition  restricts discussion on other impairments of individuals with ADHD. A little known difficulty in temporal processing is linked to vital roles of cool EFs, such as attention, inhibition, and working memory in time processing~\cite{fleming2012developmental}. The lesser known hot EFs are linked to the emotional dysregulation aspect~\cite{shaw2014emotion}, with consequences such as quick temper or a ``short fuse''; low frustration tolerance; rapid or intense mood swings and heightened emotional reactivity that are related to self-regulation.

Vohs and Baumeister~\cite{vohs2004understanding} used the analogy of a thermostat that adjusts to maintain room temperature, stating that self-regulation refers to the ability of an individual to control or adjust their own behaviours, emotions, and responses to align with certain standards or goals. This control can be both conscious and unconscious, and encompasses various aspects, including regulating thoughts, emotions, impulses, appetites, task performances, and attention.  Barkley~\cite{barkley2011important} recognizes that EF and self-regulation share a similar definition, and that behavioural inhibition is essential to their performance. His theory suggests that ADHD is primarily linked to a deficit in such inhibition~\cite{Antshel2014}. This condition causes a chain reaction of issues in other EFs, making individuals experience forgetfulness, challenges with time management, and difficulty planning for the future. This affects their ability to organise actions over time and sequence complex tasks directed towards future goals. The authors also reinforce challenges in emotional and motivational self-regulation, leading to, e.g., more impulsive emotional responses, a lack of objectivity in reacting to situations, difficulties in understanding others' perspectives due to immediate emotional reactions, and challenges in calming strong emotional responses.

\subsection{ADHD in the Workplace}
Several studies exist that focus on the performance of employees with ADHD in the workplace.
A recent study by Das et al.~\cite{das21wfh} discusses challenges of neurodivergent professionals when working from home during the pandemic.
While focusing on neurodiversity, 23 of their 36 interviewees reported having ADHD.
Furthermore, the majority of the interviewees works in software-related roles.
The study finds that common challenges are creating/maintaining an accessible workplace, negotiating accessible communication practices, and balancing productivity and well-being. 

A questionnaire-based study by Fuermaier et al.~\cite{fuermaier2021adhd} finds that people with ADHD often have difficulties meeting their own perceived potential at work, performing efficiently, or taking on new tasks.
Similarly, they find that inattention is correlated and might predict with work-related problems. 

There is also criticism in the rising trend of self-diagnosed Adult ADHD, influenced by popular media~\cite{conrad2000hyperactive}. Many adults, after encountering ADHD descriptions, identify with the symptoms and actively seek a professional diagnosis, leading to potential over-diagnosis. Skeptics raise concerns about the validity and authenticity of these diagnoses, noting that adult ADHD has become a convenient medical explanation for personal failures, such as job loss or divorce. 

Finally, a recent survey by McDowall et al.~\cite{mcdowall2023neurodiversity} among 1117 people, 127 employers and 990 neurodivergent employees, discusses challenges, strengths, and strategies in the workplace.
611 of the participants report that they have ADHD.
Among other findings, the survey finds that over 45\% of the participants with ADHD find it likely that they will leave their current job.

\subsection{Impact on Software Engineering Activities}
Literature about software engineers with ADHD is very limited. In the first study about challenges of neurodiverse software engineers~\cite{morris15}, among the 10 interviewees, only one subject had ADHD while the rest was on the autism spectrum. This person sought professional help after a negative performance review. He struggled with time management, task prioritisation, and multitasking, especially during verbal conversations. As strategies to better cope with his condition, he requested permission to audio record meetings for better task recall, written instructions for clarity, private weekly check-ins with his manager to align on tasks, and more detailed communication to avoid ambiguities. Gama and Lacerda~\cite{gama2023understanding} report a pilot study on supporting neurodivergent individuals.
They report results from interviews with four people with ASD and ADHD.
The study reports challenges such as remembering discussions from meetings, coping with sudden change, or dealing with impulsivity.

\section{Method}
We study the experiences of people with ADHD in SE, which inherently depend on the specific real-life context in which these people work.
Therefore, we used case study as a method.
We follow the guidelines by Runeson et al.~\cite{runeson2012case}, as well as the ACM Empirical Standards\footnote{\url{https://github.com/acmsigsoft/EmpiricalStandards/blob/master/docs/CaseStudy.md}}.
In the following, we discuss the method in detail.

\subsection{Study Design}
The study design we use is an exploratory, single-case design with one unit of analysis.
The case we studied is \textit{software engineers with ADHD}.
We consider the term software engineer broadly, also including people in managerial roles that have a technical software engineering background.
The unit of analysis comprises all \textit{work-related activities}, including interactions with other people at work.

The study was conducted by three researchers.
The first and third author are SE researchers experienced in qualitative methods and human factors.
The second author is a student who familiarised himself with qualitative methods during the study.

We follow a constructivist world view in which different actors construct their own truth through their perceptions of the world.
To better understand these individual perceptions, we chose to use interviews as a primary means of data collection.
Given that we used an exploratory, constructivist research approach, we started the study with no pre-defined theoretical base.

\subsection{Data Collection and Analysis}
We recruited study participants through advertisements on Twitter/X and LinkedIn, using hash tags related to ADHD and neurodiversity.
Furthermore, we encouraged participants to advertise the study through their networks.
To take part in the study, participants had to fill in an informed consent form\footnote{Included in the published dataset.} that was provided as an electronic form.
In the form, participants could give or deny consent to participating at all, to record the interview, and to having their anonymised transcript published after study completion.
They were granted the right to withdraw their consent at any time.
For publishing the transcripts, we obtained their consent a second time after we sent them the transcripts and the preliminary results.
All participants consented to participation and recording, and six people consented to publication of their transcript.
The study was conducted at universities in northern Europe and South America, where no ethics approvals were required under the local regulations.
The interviewees with their role or area of work are listed in Table~\ref{tab:interviewees}.
In terms of gender, 13 identified as men, 2 as women, 1 as a trans-woman, and 3 did not disclose their gender.
Six interviewees work in South America, five in Europe, four in North America, and two in Asia.
Two interviewees did not disclose their location.
Interviews took between 20 and 60 minutes.
\begin{table}
    \centering
    \caption{Interviewees with experience and role.}
    \label{tab:interviewees}
    \begin{tabular}{cll}
        \textbf{ID} & \textbf{Role or Area of Work} & \textbf{Experience} \\
         I1 & Information Security & 29 years \\
         I2 & QA Engineer & <1 year \\
         I3 & Security Analyst & 20 years \\
         I4 & Software Development & 2 years \\
         I5& Founder & 16 years\\
         I6& Engineering Manager & 13 years\\
         I7& Senior Software Engineer& 7 years\\
         I8& QA Engineer & 5 years\\
         I9& Mobile Development & 15 years \\
         I10 & Mobile Development & 10 years\\
         I11& Director of Engineering & >20 years\\
         I12& Senior Engineering Manager & 20 years\\
         I13& Software Development & 1 year\\
         I14& Software Development & 5 years\\
         I15& Senior Data Engineer & 7 years\\
         I16& Software Development & 2 years \\
         I17& Software Development & 3 years\\
         I18& Software Development & 1 year\\
         I19&  Senior Software Engineer & 6 years \\
    \end{tabular}
\end{table}

We collected and analysed data in an iterative fashion.
Specifically, we used in vivo coding throughout the analysis, i.e., using the interviewees' own words as codes \cite{saldana2021coding}.
In vivo coding helps to avoid misinterpretation of the interviewees' spoken words.

Coded statements were then gathered in the electronic whiteboard software Miro\footnote{\url{https://miro.com}} and iteratively grouped into related categories.
We jointly discussed the categories after 3 interviews, after 10 interviews, and then at least weekly until we reached 19 interviews.
After the first two discussions, we also adapted the interview instrument\footnote{All interview instruments are included in the published dataset.}.
After 17 interviews, we summarised the results graphically and in writing and sent them out to the interviewees for member checking.
Furthermore, to account for a limited background in Psychology among the researchers, we discussed these results with a researcher in Psychology at one of our universities.

Finally, we performed four semi-structured interviews with experienced managers (14, 16, 23, and 20 years of experience in SE; one woman and three men) recruited through personal contacts, discussing the implications of our findings for the SE industry.
These interviews took approximately 30 minutes each, starting with a presentation of our findings.
Informed consent was obtained orally, and we took notes only.
The managers' views (M1-M4) are reported in Section~\ref{sec:discussion} with a joint discussion on industrial implications.

\subsection{Validity Threats}
We present threats to validity in terms of transferability, credibility and confirmability \cite{petersen13}, as we follow a constructivist world view.

\subsubsection{Transferability}
Transferability describes to what extent study results apply to similar cases \cite{petersen13}.

We only report categories supported by at least three different interviewees, to ensure sufficient triangulation.
Similarly, we include interviewees with different roles, experience, and geographic distribution.
Therefore, we believe that the results are transferable to other people with ADHD working in the software industry. 
Nevertheless, the broad spectrum of ADHD-related symptoms and specific work context can vary substantially.
To mitigate this threat to some extent, we discuss our findings, which are based on inductive research, in relation to existing theory from ADHD research.

While most categories remained stable after analysing 10 interviews, we did not reach saturation.
We decided to stop data collection after 19 interviews, since the only newly-added categories related to symptoms more commonly associated with ASD and Dyslexia, which have a high co-morbidity with ADHD \cite{hours2022asd,mcgrath2019there}.
Additionally, we created hypothetical associations between different categories.
Validating these remains future work.

\subsubsection{Credibility}
Credibility describes whether findings are reported truthfully, or have been distorted by the researchers \cite{petersen13}.

We used recordings and verbatim transcripts in our analyses to avoid distortion.
During the analysis, we maintained a chain of evidence using in vivo codes, memoing, and frequent discussions among the three authors.
Similarly, we use extensive quotes when reporting our results.
Finally, to increase credibility, we publish all versions of our interview instruments, and six anonymised transcripts in our dataset.

\subsubsection{Confirmability}
Confirmability describes the extent to which researchers' conclusions follow from the observed data \cite{petersen13}.

Our main means to ensure confirmability is the use of in vivo codes that are quoted in the results, combined with member checking among all interviewees.
The preliminary results section, together with a mapping of interviewees to categories was sent to the interviewees with a one-week period to give feedback.
Several interviewees replied with comments, but without any corrections to the reported categories. 
\section{Results}
In the following sub-sections, we report our findings according to challenges, strengths, and strategies reported by our interviewees.

\subsection{Challenges}
\begin{figure}[ht]
  \centering
  \includegraphics[width=\linewidth]{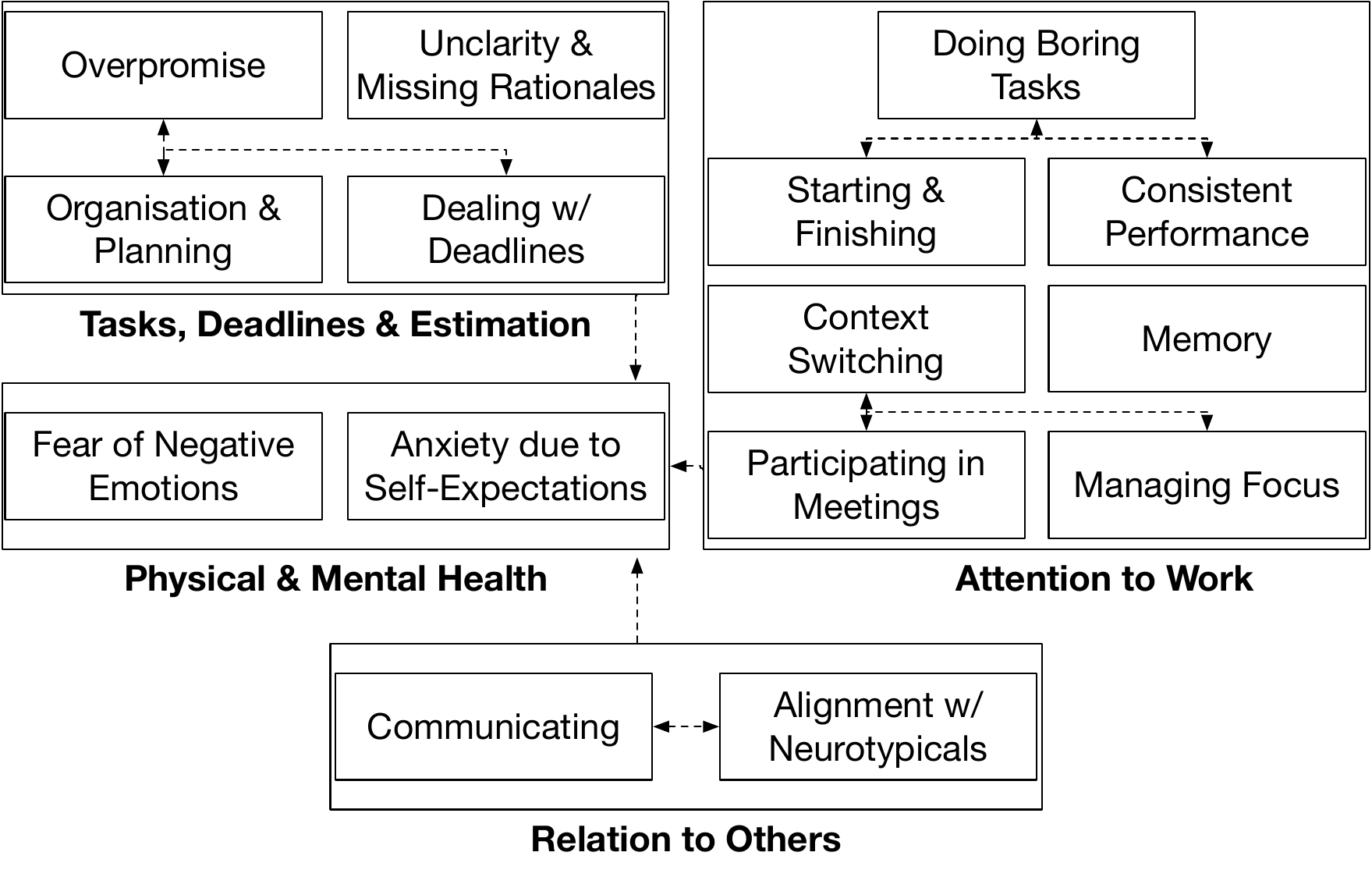}
  \caption{Challenge Categories. Arrows between categories depict hypothetical associations.}
  \Description{15 Challenge categories grouped into 4 higher-level categories. Arrows between categories depict hypothetical associations.}
  \label{fig:challenges}
\end{figure}

According to our results, people with ADHD deal with various different challenges (see Figure~\ref{fig:challenges}).
Below, we report these according to the four main categories.

\subsubsection{Tasks, Deadlines \& Estimation}
Three challenges relate to organisation of various tasks and activities, related deadlines, and estimation of necessary time.

\paragraph{Over-Promise} 
Three interviewees report the tendency to over-promise on their deliveries.
For instance, two interviewees state that they would commit to too many projects:
\textquote[I8]{I take on too many projects.}
\textquote[I6]{I will say yes to everything.} 
This over-commitment then leads to overtime.
\textquote[I6]{[I would] work 16 hours a day, and also working weekends}

\paragraph{Organization \& Planning} Eleven interviewees state difficulties with organisation and planning.
The statements range from general lack of organisation skills to more specific issues.
For instance, two interviewees raise that they have difficulties estimating time.
\textquote[I6]{I have no sense of time.}
\textquote[I12]{I just don't really have a good sense of, like, how long something will take.}
    
In others, the tendency is rather to either start things without planning sufficiently, or, vice versa, over-planning activities.
\textquote[I8]{I usually am the person that just like dives right in and starts everything off before actually like taking the time to plan anything very thoroughly.} 
\textquote[I9]{I tend to sometimes over-plan stuff.}

\paragraph{Dealing with Deadlines} Deadlines are challenging for several of our interviewees.
In some cases, this relates to difficulties with estimating time, as mentioned above.
However, in many cases the difficulties rather relate to actively working towards deadlines instead of prioritising other tasks.
\textquote[I5]{Had a task for this for like 3 months in my to do list [..] 72 days overdue, and I wouldn't do anything about it}
Similarly, several interviewees state that they would have issues remembering deadlines.
\textquote[I3]{[..] If I can't see it, I'm not going to remember it.} 
Finally, the fact that deadlines exist cause stress and anxiety in some interviewees.
\textquote[I9]{[I am] quite stressed by deadlines.}

\paragraph{Unclarity and Missing Rationales} 
Six interviewees state that they find a lack of clarity challenging.
This relates both to the way of working at the company, and to goals of development tasks.
\textquote[I12]{The biggest issue for me was not having a clearer 'why'.} 
\textquote[I9]{I struggle with this navigating structures and finding who to talk about.}

\subsubsection{Attention to Work}
Seven challenges emerged in relation to paying attention and maintaining focus to various work tasks.

\paragraph{Doing Boring Tasks} As a result of preferring instant gratification, performing tasks that are perceived as mundane or not interesting is difficult for 14 interviewees.
\textquote[I5]{It's not necessarily a specific area. It's more about what feels fun now versus what isn't.} 
\textquote[I12]{I just want to get to the interesting part}

A few interviewees describe more specific areas that tend to be tedious.
Two people mention writing repetitive code or boilerplate code.
\textquote[I4]{Repetitive tasks [..] something which [..] requires a lot of coding to set up, that is the most challenging part.}

Three interviewees state that non-technical parts like planning projects or management tasks are hard to complete:
\textquote[I9]{Where it's less technical, it's more business and this kind of stuff, I struggle [..].}
\textquote[I10]{the non-software part of engineering [is boring].}

Finally, four interviewees state that they have difficulties performing tasks that do not provide instant reward, which is related to delay aversion.
\textquote[I1]{[I struggle with] tasks where I'm not getting dopamine hits.}
If this is not the case, focusing on the task is difficult
    \textquote[I12]{[..] keeping, like, the final thing, the final reason that I'm doing a thing, the final goal, on top.}
\paragraph{Memory}
Three interviewees state that they have difficulties remembering things, especially short-term memory.
\textquote[I13]{My memory is sort of contained in the short term}
\textquote[I7]{I've noticed that my ability to remember things has got really bad, especially like within, like a 5-minute period}

\paragraph{Starting and Finishing} Starting and finishing tasks is considered difficult by eleven interviewees.
For some, this challenge relates to a lack of motivation.
\textquote[I3]{[..] lack of motivation and it just being able to really focus and dive into things, it's just not there.}
\textquote[I13]{When I feel I'm near to that goal, or I actually reach that goal, I wouldn't go any further than that.}
However, others mentioned that they instead perceive this challenge like a block.
    \textquote[I12]{the first 5 percent are impossible. And then last 15 percent is like torture.}
    \textquote[I10]{the idea of like time paralysis, when, like you have a meeting in an hour so like you won't start anything, because, you know, you won't be able to get into the... that mindset.}

\paragraph{Consistent Performance} 
Various challenges with organisation, planning, as well as with keeping focus over a longer period of time lead to issues maintaining consistent performance at work, as stated by four interviewees.
\textquote[I1]{[I have a] fairly narrow window [..] where I am able to execute at a consistent level}
Specifically, one interviewee described their productivty as 'spiky'.
\textquote[I10]{Over like a long enough time period [..] output is the same, or higher or whatever [..] but mine is very spiky.}

\paragraph{Context Switching} 
Linked with focus troubles, interviewees shared difficulties switching context.
\textquote[I9]{Switching to [a] state, that takes me quite a bit of time.}
While context switching is difficult for most people, individuals with ADHD might have an even harder time.
For instance, one interviewee shared:
\textquote[I1]{It can take me almost an hour to get the entire structure and everything like loaded into my mind, and then somebody will come by me like, 'Hey, did you see the football game?'.}

\paragraph{Maintaining Focus} Dealing with distractions and maintaining focus on a task is mentioned by eleven interviewees.
In general, this relates to easily getting distracted by things that capture the interest of the interviewees more than the tasks at hand.
\textquote[I3]{Shiny objects, and I'll go in 10 different directions.}
\textquote[I4]{People with ADHD, concentrating for 1.5 hours, is really the first battle to fight.}
However, the work environment in particular provides a common source of distraction.
    \textquote[I1]{For someone with ADHD, [..open office] is just a recipe for disaster }
\textquote[I14]{I couldn't listen to music, and I couldn't concentrate on anything because it was so noisy.}

\paragraph{Participating in Meetings} Nine interviewees state that it is hard for them to participate in meetings.
Specifically, they mention that they get very easily distracted so that they cannot follow the meeting.
\textquote[I1]{I wind up like doodling. I'll say things like, 'Oh. [..] Sorry the zoom glitched out.'.}
\textquote[I10]{Someone would get on a very boring tangent and I would just, like, start scrolling on my phone.} 

Additionally, several of the interviewees express that they feel meetings are frequently a waste of time.
\textquote[I5]{Meetings are a waste of time. If nobody's prepared, that means we gonna have a meeting that will result in another meeting}

\subsubsection{Relation to Others}
As software is primarily engineered in teams, the relation to other engineers is important.
However, several challenges arise in this context.

\paragraph{Alignment with Neurotypicals} 
Ten subjects describe challenges aligning with neurotypical colleagues.
They describe that others sometimes struggle to follow their line of thought.
\textquote[I1]{Sometimes I couldn't get them to see [..] That's where I usually left.}
\textquote[I6]{I will spend like a few hours just to explain to people.}

Similarly, several interviewees have the feeling that they are not in the right place, and should behave differently.
\textquote[I6]{[I'm] not playing the role that I'm supposed to be.}
\textquote[I2]{[..] I put a lot of pressure on myself to be normal as much as I can.}
This aspect also relates to the fear of stigmatisation.   
    \textquote[I4]{The main challenge is the stigma.}
\textquote[I3]{Job applications. They ask you if you suffer from any of these things [..] and I don't even like answering those. You know there is that stigma around it.}

\paragraph{Communicating} 
As a third category, we find that seven interviewees struggle with communication style.
In some of these cases, the communication style differs, and is considered rude or confusing by others.
\textquote[I14]{I was so wired and agitated that the guys from another team were saying I was being rude.} 
    \textquote[I5]{some things I just say unfiltered.}

In other cases, interviewees state that they have a hard time directing others, e.g., to work overtime.
\textquote[I16]{I am absolutely terrible at [pushing others to put in extra work, overtime].}

Interestingly, one person states it is easier to express thoughts in code.
\textquote[I11]{Somehow, in code it's easier, but in words, it's harder.}

\subsubsection{Health}
Several health-related challenges were raised by our interviewees.

On a general level, the accumulated experiences of people lead to difficulties in maintaining a good physical and mental health.
The interviewees raise issues ranging from depression and trouble sleeping up to substance abuse.
\textquote[I6]{Depression, anxiety, substance abuse.}
\textquote[I8]{Anxiety and depression could be because [of] undiagnosed ADHD.}

    \textquote[I3]{I remember having those days that you know, just could not get anything done. And I would come home just frustrated because I knew I wasted a day[..]}

\paragraph{Fear of Negative Emotions} 
Four interviewees a fear of negative emotions.
For two, this relates to fear of rejection.
\textquote[I6]{[I have] rejection sensitivity dysphoria, [..] fear of rejection. [..] I cannot handle people saying no.}
    
As several of our interviewees work in managing positions, this is also an issue that arises when the performance of team members is not satisfactory.
\textquote[I11]{It's the absolute worst part of my job. I wish every single person could just do great.}
    
\paragraph{Anxiety due to Self-Expectations} 
Most interviewees have experienced difficulties at work that can be attributed to ADHD, e.g., due to issues with meeting deadlines.
This results in anxiety, as stated by six interviewees.
    \textquote[I13]{when I face box or issues that I can't resolve immediately I tend to get very anxious.}
In some cases, this relates to their own set expectations:
    \textquote[I8]{I am always my own hardest critique.}
\textquote[I12]{I had a lot of impostor syndrome, so like learning new systems I thought was difficult.}

 \subsection{Strengths}
\begin{figure}[ht]
  \centering
  \includegraphics[width=\linewidth]{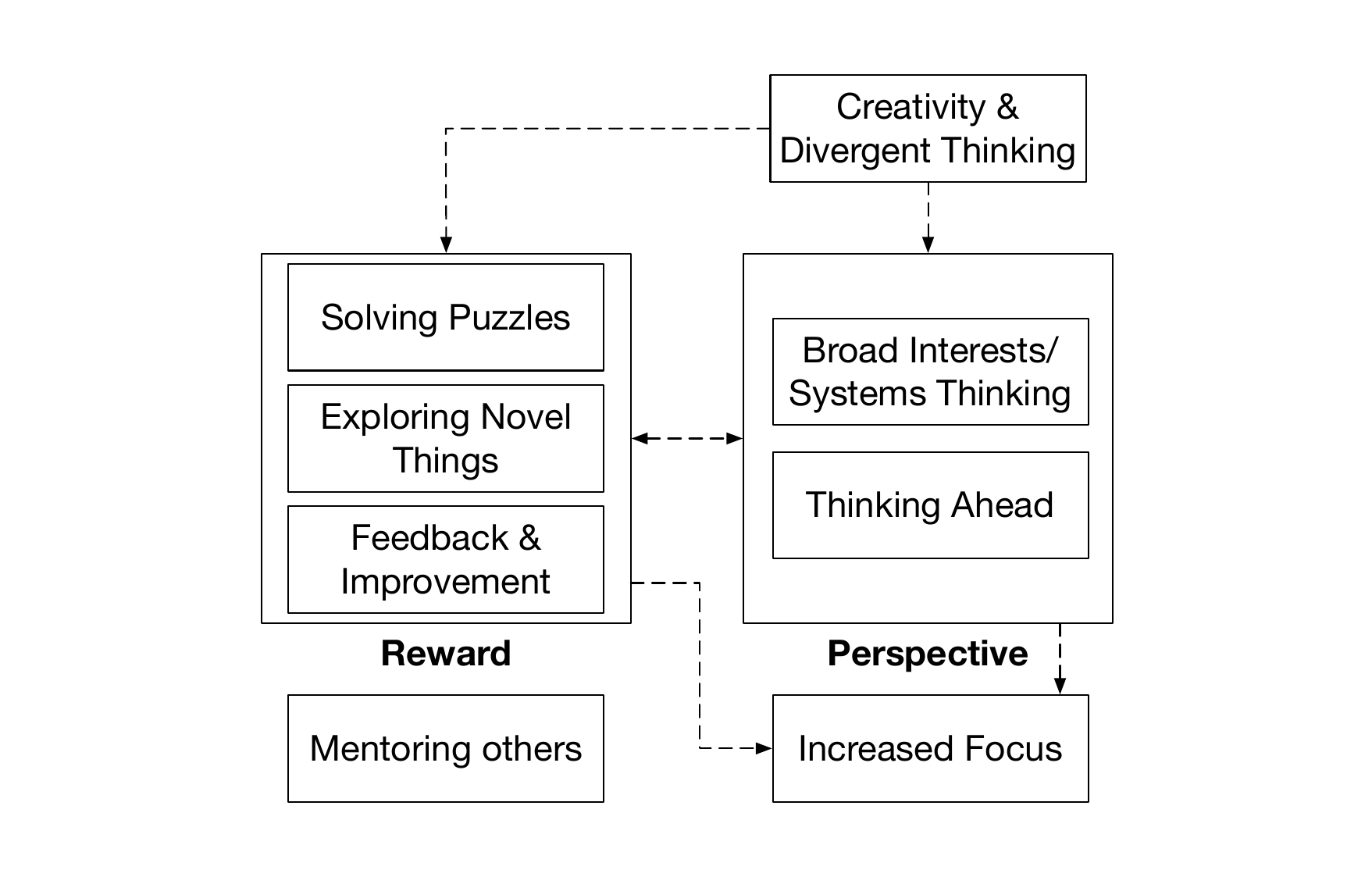}
  \caption{Strengths reported by the interviewees. Arrows between categories depict hypothetical associations.}
  \Description{Eight Strengths reported by the interviewees. Arrows between categories depict hypothetical associations.}
  \label{fig:strengths}
\end{figure}

While challenges are common, the interviewees also raise numerous perceived strengths (see Figure~\ref{fig:strengths}).
We discuss them below.

\subsubsection{Creativity \& Divergent Thinking} Twelve interviewees state that they have the ability to be creative, essentially being good 'out-of-the-box' thinkers.
[..]  \textquote[I4]{[I'm] always coming up with creative ideas.}
\textquote[I12]{I'm definitely better at like that kind of yeah, creative, generative conversations.}

However, some specify that creativity is primarily linked to the various interests they have.
\textquote[I5]{Integrating a lot of those things that you dipped your toes into. [..]  I think that's what helps with this divergent thinking, right?}  

Finally, two interviewees also describe that they do not consider what they produce creative, even if it is novel:
\textquote[I11]{I'm creative but I don't feel like it's... when most people say creative it's what they think of.}
\textquote[I7]{I've just like glued a few things together and built this thing that does a lot of stuff that no one else has done.}

\subsubsection{Rewards}
Three strengths relate to tasks that provide the feeling of having a reward.

\paragraph{Solving Puzzles} 
Seven interviewees state that they are good at solving puzzles.
They state that they are good at thinking through multiple different steps.
\textquote[I11]{Some background process in my brain is putting all these puzzle pieces together and figuring that out.}
    \textquote[I12]{I'm better at making connections between things.}

This strength specifically relates to software engineering activities like investigating incidents or fixing bugs.
\textquote[I3]{Really good at investigating security incidents (big puzzles).[..] It branches out into 10 different directions, and you never know which thread to pull on.}
\textquote[I11]{I'm pretty good at seeing like 'Aha! Like this system that we're using. It is broken because of this.'}

\paragraph{Exploring Novel Things} 
Connecting different pieces of information was mentioned by interviewees related to solving puzzles (see above).
However, it also leads to the strength to explore novel things, as reported by seven interviewees.
\textquote[I1]{I like the novelty [..] always exploring.}
\textquote[I11]{I guess I'm good at picking up new tech things.}  
In turn, this strength may encourage them to share novelty at work.
\textquote[I2]{I've been taking the initiative at work, like bringing in new technology to my team.}

\paragraph{Feedback and Improvement Tasks} Seven interviewees state that they are good at tasks that provide immediate feedback, or tasks that consist of iteratively improving something.
\textquote[I1]{it doesn't even have to be a reward, it just has to be feedback}
This is why they enjoy doing debugging and also the product side of a project.
\textquote[I11]{I love refactoring and debugging.}

\subsubsection{Perspective}
Two strengths relate to a broad perspective on software systems and their evolution.

\paragraph{Broad Interests/System Thinking} Twelve interviewees state that their interests are very broad.
They consider this a strength, as it enables them to think broadly and consider various aspects of a system and its environment.
\textquote[I1]{I've got such a broad interest in various aspects.}
\textquote[I8]{I've always just been very... in everything, doing everything.}
Furthermore, it allows them to make connections that other people are less likely to see.
\textquote[I6]{[I'm] good at synthesizing all these different requirements.}
\textquote[I5]{your interest switches. So you kind of have this bigger picture view.}
\paragraph{Thinking Ahead} 
Six interviewees state that they see a strength in the ability to think ahead.
    \textquote[I2]{I just told the tech lead [..] they're doing this like this. [..] they're already thinking about this.}
\textquote[I10]{I go a little deeper than what a lot of people do.}
Due to this ability, they have an improved ability to see how systems or projects will evolve in the future.
\textquote[I5]{How the system will evolve in terms of where we are now. [..] like a product road map type of thing.}

\subsubsection{Increased Focus} 
Seven interviewees state that they can sometimes be extremely focused and productive on parts of a project.
\textquote[I1]{Work-related [increased focus] [..] can be incredibly helpful}
\textquote[I10]{it's interesting to me. So I get a kind of a hyper focus on it.}
However, the interviewees also state that this increased focus can be both very draining and rare to achieve.
\textquote[I5]{If it really clicks you just you go on loop. [..it] is draining. But in this makes you way more, let's say, forward looking.}

\subsubsection{Mentoring/Training Others} As a final strength, mentoring colleagues is a strength six interviewees stated as well.
\textquote[I3]{I enjoy like training other people like the lower level analysts that I work with.}
\textquote[I11]{I love... you know, employee development and employee growth.}  \subsection{Strategies}
\begin{figure}[ht]
  \centering
  \includegraphics[width=\linewidth]{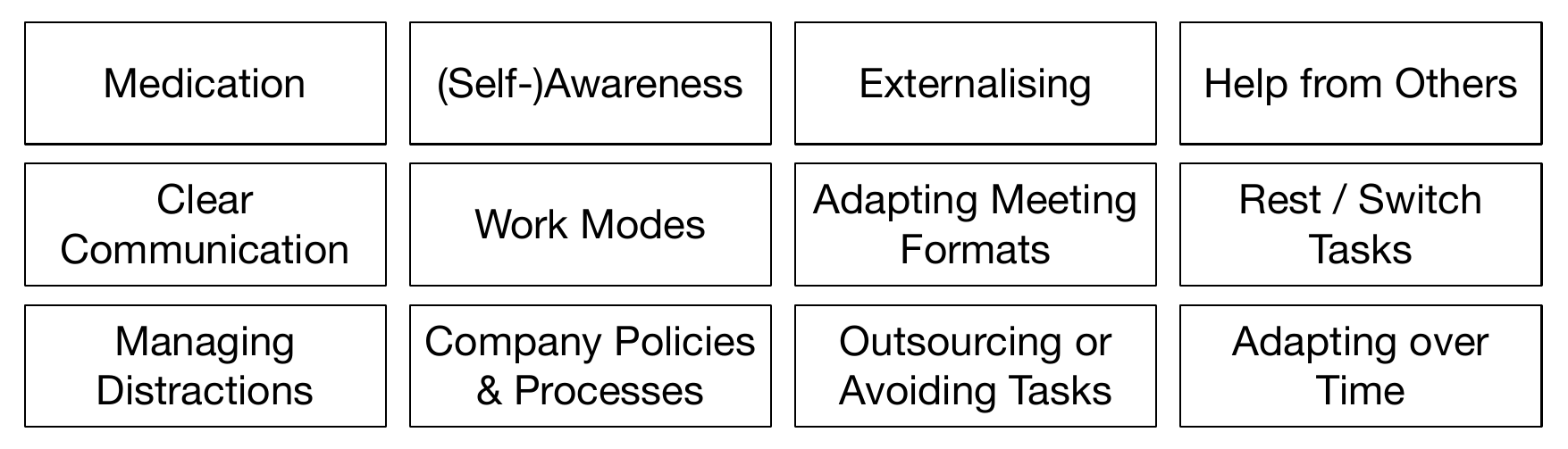}
  \caption{Reported Strategies.}
  \Description{Twelve reported strategies.}
  \label{fig:strategies}
\end{figure}

As a final part of our results, we elicited twelve strategies (see Figure~\ref{fig:strategies}) the interviewees use to overcome challenges, or to better use their strengths.
Below, we report these strategies.

\paragraph{Medication} Medication plays an important role for almost all of our interviewees, and helps them significantly.
\textquote[I3]{Medication helps a lot.}
\textquote[I6]{it is helping me a lot compared to [..] before.}
However, a few interviewees also state that the side effects can be difficult, which is why they try to avoid medication.
\textquote[I13]{Partly because of side effects. But I also want to try not to rely on medication that much.}
\textquote[I8]{I felt I just could not feel happy at all while I was on that.}

\paragraph{(Self-)Awareness} 
Eight interviewees state that being aware that they have ADHD helps them a lot to understand and to adapt better.
\textquote[I5]{I could have been way more effective if I knew what was going on.}
\textquote[I13]{I understand that I'm going through this and I'll feel this way so I can respond in a more appropriate way.}

Daily reflection and being patient are two ways raised by the interviewees to further improve their self-awareness.
\textquote[I10]{Being a little bit patient, and open to like understanding that.}
\textquote[I13]{It helps a lot, which is daily reflection.} 

\paragraph{Externalising} 
Externalising things in different ways is a strategy mentioned by eleven interviewees.
The most popular means, todo lists, help with keeping track of tasks and activities, remembering deadlines, and staying productive.
\textquote[I5]{Now it's on my to do list. I'm gonna execute on it.}
\textquote[I12]{My lists are so all over the place, and I'm actually getting a shit more done.}
Some interviewees intentionally mix the kind of items on their lists, e.g., including both work-related and personal items.
\textquote[I3]{Listing of projects, all short term projects and goals [..] not all work related.}

As support tools, several interviewees use calendar apps.
However, physical lists and sticky notes are mentioned as well.
\textquote[I12]{I used to try to use like todo list or you know, these like very structured to do list apps.}
\textquote[I3]{Even now I've got a a small sheet of paper behind me, or in front of me. }

In addition of keeping track of todos, several interviewees take notes of all information they consider useful in order to remember details in the future.
\textquote[I11]{I have a document per report where I'll just keep information about them [..]  I also put dates on for things because I made mistake like 'Oh, how was your vacation?', they're like 'that was 6 months ago.'}
\textquote[I7]{The blog posts that I do is definitely a thing of it, it is a way for me to remember things.}

Since our interviewees typically have a programming background, four interviewees stated that they build their own support systems that help them to deal with various challenges.
\textquote[I5]{Systems that I built to like, how I manage the team, how I communicate with them were because I have ADHD}
\textquote[I15]{I have to automate tests in everything, linters, have some minimum automation in the IDE to try to catch possible typos.}

\paragraph{Help from Others} Ten interviewees ask for help from colleagues or other people around them.
Four interviewees mention therapy specifically.
\textquote[I4]{[..] And so therapies could help.}
\textquote[I16]{Therapy [Cognitive Behavioural Therapy (CBT)] was kind of liberating for helping me find my functional self beyond the medication.}

Another method that was frequently mentioned is 'body doubling', where you work with someone else in the same room, even if they are not talking or working on the same thing.
\textquote[I13]{I do like soft Co-studying with my friends as well.}
\textquote[I12]{Also body doubling is so huge for me.}
One of the interviewees mentioned specifically pair programming in this context:
\textquote[I19]{But I think it [Pair Programming] can help someone with ADHD because if they get distracted, there's someone there to pull them back in}

Coaching is mentioned by two people, and having an accountability buddy by one.
\textquote[I14]{The leader made himself available, he saw that I was technically limited.}
\textquote[I17]{It's 'cause when I make plans with someone, I'm committing to them.}

\paragraph{Clear Communication} 
Seven interviewees report that they use clear communication to make others understand how they are thinking and which challenges they might experience.
\textquote[I11]{Having this negotiation of how you communicate, I find, this very helpful.}
\textquote[I11]{'Tell me exactly how you want us to interact'. And like, 'Oh, my God, it's so great'. Just..it's clarity.}

Finally, one interviewee stated that several colleagues got diagnosed due to them talking about their symptoms.
\textquote[I7]{When I got my diagnosis and I was very vocal with my team about it, that then also led to a few of my colleagues getting diagnosed.}

\paragraph{Work Modes} 
Ten interviewees suggest modifying different work modes.

Six interviewees state that it is helpful to work from home.
\textquote[I3]{[..] Working from home's been really helpful, too.}

One of the reasons why this strategy helps, is due to the control over the environment.
    \textquote[I12]{Having a lot of control over my environment helps.}
\textquote[I3]{[I] don't have people looking over my shoulder every minute of the day.}

Similarly, having flexible work hours is reported as helpful by seven interviewees.
\textquote[I10]{I'm fairly nocturnal person. So for me, working like from 9 to 5 is not always the most conductive to be like a good employee.}

Finally, one person suggests that smaller sub-groups within a company helps them.
\textquote[I9]{More autonomous kind of... subgroups.}
    
\paragraph{Adapting Meeting Formats} 
Since meetings can be challenging for many interviewees, they work on adapting meeting formats to make them easier to follow, or reduce the meeting time.
\textquote[I2]{[..] we definitely need these meetings, but they could be shorter.} 
\textquote[I1]{Either be like a working meeting or a quick, like a standing meeting.}

One frequently-mentioned aspect is that agendas help preparing for and following a meeting.
\textquote[I11]{Meetings need agendas.}

Finally, two interviewees shared ways of better keeping attention during remote meetings.
\textquote[I10]{For a couple of meetings like I will have like a fidget cube.}

\paragraph{Rest / Switch Tasks} 
Another strategy developed by three interviewees is to accept when they are stuck on one task, and take some time to rest and do something else.
\textquote[I10]{Being more kind with yourself  [..]  it's fine to have a day off.}
\textquote[I11]{I shouldn't go look at a Youtube video, but if I look at a pull request, it's work. So it's like doing a distracting thing but that's work, so that actually helps.}

\paragraph{Managing Distractions} 
Since many of our interviewees are easily distracted, removing distractions in their environment can help.
    \textquote[I10]{Not take my laptop to the meeting, because I would otherwise... and like to leave my phone behind.}

Also, some people talk about having 'good' distractions that help them focus.
\textquote[I12]{Music and sort of different kinds of like beats and rhythms [..are] really helpful.}
\textquote[I3]{I got the TV on with something playing in the background, just to get some background noise.}
    
\paragraph{Company Policies and Processes} The way companies work play an essential role for our interviewees.
They mention various aspects that could help them perform better.
First, interviewees mention diversity policies and generally talking about neurodiversity.
\textquote[I11]{I had to find a place where people cared less about the norms because I really didn't fit in in a lot of places.}
\textquote[I8]{Work environment should just embrace how people are and try to work with it.}
This is also raised in relation to educating neurotypical people with respect to ADHD.
\textquote[I13]{Companies could do more in terms of educating line managers.} 

A second aspect is to offer flexibility with respect to deadlines and to the work environment, e.g., tooling.
\textquote[I6]{View the idea of deadlines differently.}
\textquote[I16]{And not having that pressure to clock in, especially being on time all the time.} 
\textquote[I5]{Some companies force everybody to use the same editor, and I'm like, 'I need my Emacs'.}

Several interviewees further mention that an executive coach or an assistant could help them.
    \textquote[I12]{They should offer them access to an executive assistant or a ADHD coach.}

\paragraph{Outsourcing or Avoiding} Related to offering an assistant, as stated in the previous paragraph, interviewees tend to outsource or avoid tasks that they find difficult.
\textquote[I6]{[I do] not do that part of the job [time estimation] by myself.}

    Some highlighted how important it is to switch roles when it does not suit you.
\textquote[I5]{Find the job that suits you [..] not every role will be suitable.}
\textquote[I8]{Using your strengths where it's needed instead of sticking you on a project that you're not functioning well on.}

\paragraph{Adapting over Time} Several of our interviewees were diagnosed with ADHD at adult age.
Therefore, they had to deal with their challenges over a longer period time without being aware of ADHD.
As a result, they adapted over time and practiced to deal with their challenges.
    \textquote[I5]{I had to work around so many things already.}

Some of them practice real situations at home to be able to do those things at work.
\textquote[I11]{I practice on my cats. I do, I need to practice it because I get so emotional, and it's not the other person.}

\section{Discussion}
\label{sec:discussion}
In the following, we discuss our findings in relation to the clinical perspective on ADHD.
Furthermore, we discuss the implications of our findings on SE industrial practice.

\subsection{Clinical Perspective}
\paragraph{Challenges} Most of the challenges we identified can be mapped to relevant evidence from literature specialised on ADHD. In the category named \textit{Tasks, Deadlines \& Estimation}, the challenge of over-promising is a consequence of impulsivity, which causes a worker to over-commit or frequently switch tasks without completing~\cite{nadeau2005career}, thus also affecting \textit{Dealing with Deadlines}.
\textit{Organization \& Planning} and \textit{Dealing with Deadlines} are clearly related to \textit{planning}, which is one of the cognitive processes that comprise cool EFs~\cite{Goldstein2014}, which may lead to
procrastination, inadequate planning and missed
deadlines in people with ADHD~\cite{langberg2008organizational}. The temporal aspect of adults having difficulties in time discrimination, resulting from inhibitory control problems, can also be related to deadlines~\cite{valko2010time}. \textit{Unclarity and Missing Rationales} are related to the difficulty of following through on instructions, especially when ambiguity may lead to exacerbating time to understand and act~\cite{barkley2010adhd}. There are many distractions to both neurotypical and neurodivergent people, due to excessive usage of smartphones, social networks and technology in general~\cite{ward2017brain}. However, as emerged in \textit{Attention to Work}, people with ADHD typically struggle with that. Challenges such as \textit{Doing Boring Tasks} or \textit{Starting and Finishing} are related do delayed aversion and the search for quick rewards~\cite{fleming2012developmental,crone2004developmental} but also difficulties regulating response inhibition for stopping tasks~\cite{nigg2005executive}. The rare SE literature~\cite{gama2023understanding} on neurodiversity brings some evidence of ADHD developers tending to do over-engineering in tasks they enjoy and have trouble stopping. Weaknesses on the working memory EF are reflected on what emerged as \textit{Memory}, but are also related to \textit{Context Switching} and remembering what was being done~\cite{wu2006attention}. 
\textit{Maintaining Focus} is an impairment directly linked to inhibitory control~\cite{barkley2006attention,barkley2010adhd}. Overall, issues with regulating attention can range from extreme hyperfocus to complete lack of concentration, and can lead to other challenges like trouble switching between tasks~\cite{hupfeld2019living}.

Concerning \textit{Communication}, literature highlights communication impairments in many people with ADHD who have different processing speed of social cues and are frequently misunderstood~\cite{nijmeijer2008attention}. In many situations, due to fear of stigma and social exclusion or rejection, neurodivergent individuals try to mask their true selves in various social settings, and keep trying to fit in~\cite{beaton2020self}. The challenges related to \textit{Health} are mostly related to emotional dysregulation~\cite{shaw2014emotion}. The term rejection sensitivity dysphoria is starting to be used to describe the greater sensitivity that people with ADHD have to rejection, causing them to set higher standards for themselves, resulting in negative feelings such as embarrassment or low self-esteem~\cite{bedrossian2021understand}.
Also, issues related to psychoactive substance abuse were linked to the impulsivity aspect of ADHD~\cite{wilens1998does}.

\paragraph{Strengths} Much of ADHD literature focuses on impairments and treatment, with limited research on its positive sides. However, some reports and studies ~\cite{white2006uninhibited,sedgwick2019positive,mcdowall2023neurodiversity} highlight attributes like heightened \textit{creativity and divergent thinking}, as well as  curiosity. Many interviewees said they were creative and good at coming up with new ideas, matching with studies that state ADHD being more creative~\cite{white2006uninhibited,sedgwick2019positive}.  There are also studies~\cite{hupfeld2019living,mahdi2017international,nadeau2005career} emphasising other positive aspects such as resilience, energy, hyperfocus, and empathy. Many of our participants like \textit{mentoring} their colleagues. This can be linked to increased empathy. Some interviewees said they do well when getting immediate feedback on their work. This relates to studies that tell us people with ADHD look for quick \textit{rewards}~\cite{fleming2012developmental,crone2004developmental}.

Empirical studies~\cite{kosaka2019symptoms,keezer2021masking} report that many people with ADHD have an above average IQ, despite their cognitive and emotional impairments. 
Many of our interviewees are also good at \textit{solving puzzles} and seeing connections, which may suggest a potential relation to higher IQ. Many participants reported liking to plan for the future and think about the bigger picture. This goes against the usual idea that people with ADHD only think about the ``now'' and cannot plan ahead. Although ADHD is linked with attention issues, some of our interviewees said that they can have many situations of \textit{increased focus} when compared to  neurotypical peers under certain circumstances, which is confirmed in ADHD literature~\cite{hupfeld2019living}.

Some people with ADHD manifest twice-exceptionality, which consists of ``demonstrate[ing] the potential for high achievement or creative productivity in one or more domains such as math, science, technology, social arts, visual, spatial, or performing arts or other areas of human productivity AND manifest[ing] one or more disabilities''~\cite{reis2014operational}. Such twice-exceptionality can mask cognitive areas in which these individuals have difficulties.
Sometimes, this means that no diagnosis during childhood and even during adulthood is sought~\cite{atmaca2022two}.  This type of detailed investigation is not mandatory for ADHD diagnosis and is financially expensive~\cite{nadeau2005career}. Participant I17 was the only one reporting that condition: \textquote[I17]{In the neuropsychological test, I found out I have this thing I didn't even know about, called twice-exceptional. It's when you have ADHD combined with giftedness.} However, we cannot conclude that other interviewees had that condition or that only twice-exceptional individuals have good work performance among people with ADHD.

\paragraph{Strategies}
Many of our interviewees have developed personal coping strategies over time. This mirrors the literature's overarching theme of targeting various ADHD aspects, suggesting that adapting strategies over time is vital for managing ADHD. Backed by empirical evidence from literature, Fleming and McMahon~\cite{fleming2012developmental} list methods and strategies to assist adults with ADHD, addressing multiple facets of the condition. These include strategies like improving executive control through mindfulness practice and externalising higher-level executive processes. For those facing challenges with motivation towards long-term rewards, the recommendation encompasses breaking tasks into smaller chunks, using contingent self-reinforcement, and practicing mindfulness. Emphasis is also placed on time management, concentration, and test-taking strategies. The guide underscores the importance of maintaining a balanced lifestyle, encompassing sleep, diet, and exercise, while also offering guidance on medication adherence and managing addictive behaviors. Finally, it touches upon the significance of emotional regulation, nurturing social relationships, and harnessing external sources of support.
We found many occurrences of such strategies being used by our interviewees. \textit{Medication} was explicitly cited as being important according to most of our interviewees. Cognitive-behavioral therapy (CBT) was cited under \textit{Help from Others}, and its effectiveness in reducing ADHD symptoms was confirmed in a recent systematic literature review~\cite{young2020efficacy}. Another important strategy that we labeled as \textit{self-awareness} consists of psychoeducation, which is learning about what causes the mental condition the person. In the case of ADHD, a self-aware (i.e., psychoeducated) person should see it as a brain difference that has both challenges and benefits.

Some of the strategies were more general, while other were adapted to the SE context. For instance, many of our interviewees use to-do lists or take notes to \textit{externalise} information, helping them staying productive. This can be linked what Fleming and McMahon referred to as externalising higher-level executive processes to improve executive control. This strategy is typically suggested in CBT and fits well as a practice for self-organisation in a software project context. \textit{Help from Others} can be simply doing a task in the presence of another person (i.e., ``body-doubling'')~\cite{mcdowall2023neurodiversity} which also appeared in the form of pair programming. Accountability partner, who could be a leader or someone from the team who helps one to monitor goal progress~\cite{young2019user} was also mentioned. This aligns with what Fleming and McMahon mapped as Establishing External Reinforcement Contingencies, which involves a support person. Some of our interviewees stressed the need to manage distractions, which aligns closely with reducing distracting stimuli and amplifying relevant stimuli~\cite{fleming2012developmental}. When exploring different possibilities of\textit{ work modes}, the preference to work from home and having control over one's environment in our results can be tied to what Fleming and McMahon reported as managing inattentive/disorganised behavior and temporal processing deficits (e.g., adding time reminders), which relates to what we called \textit{managing distractions}. Meetings are a problem mentioned in the two existing studies on neurodiversity in SE~\cite{gama2023understanding,morris15}. The \textit{adaptation of meeting formats} brings strategies to generate minutes or clearer agendas. The strategy of adaptations on \textit{companies policies and processes}, such as the mention to offering an ADHD coach, have strong empirical evidence to be effective to enhance the quality of life in people with ADHD~\cite{fleming2012developmental}.

\subsection{Software Engineering Perspective}
Our findings align well with general challenges reported in existing research on ADHD in the workplace, e.g., in \cite{mcdowall2023neurodiversity}.
However, the challenges and strengths we found also strongly relate to existing SE practice.
All four challenges categorised under \textit{Tasks, Deadlines and Estimation} are core issues in SE, especially in agile processes where regular planning, estimation, and uncertainty are key elements of everyday work.
Similarly, due to the strong team focus, \textit{Relation to Others} are central.
The quick development pace might further clash with several challenges in \textit{Attention to Work}, e.g., when struggling with \textit{Consistent Performance}, \textit{Starting and Finishing}, or \textit{Managing Focus}.
Finally, \textit{Doing Boring Tasks} and \textit{Participating in Meetings} are common requirements in SE.

Two of the interviewed managers (M1, M4) note that several challenges might not be limited to people with ADHD, e.g., anxiety (M4).
Indeed, our current analysis does not allow the conclusion that the found challenges, strengths and strategies are specific to people with ADHD.
However, two managers (M3, M4) remembered employees with ADHD who exhibited several of the reported challenges, e.g., \textit{Consistent Performance}, \textit{Communication}, and \textit{Dealing with Deadlines}

Interestingly, two managers (M1, M2) thought that many of the reported challenges could effectively be addressed through existing agile practices.
In mature \textit{agile teams}, \textit{Organisation and Planning} is done jointly, e.g., through the \textit{sprint planning} and \textit{planning poker} practices.
Issues in \textit{Consistent Performance} are addressed through mutual trust, as the team learns to deal with productivity swings.
Similarly, \textit{Relation to Others} will over time be addressed to some extent, as the team matures.
Numerous practices such as \textit{short sprints}, \textit{burndown charts}, or \textit{continuous integration and deployment} provide regular \textit{Reward} and can help addressing \textit{Doing Boring Tasks} or \textit{Managing Focus}.

However, M1, M2, and M3 all thought that the reality of software development is often different from the idealised agile development picture, which aligns with existing studies, e.g., \cite{gregory2015agile,hoda2012developing}.
Specifically, it is common that the cross-functional team practice is not used \cite{diebold2014agile} or that a command-and-control leadership style is used \cite{gregory2015agile}.
Similarly, challenging sprint deadlines are common \cite{hoda2012developing}.
Another aspect, raised by M3, is that working in service or contracting companies could be challenging for people with ADHD, as those companies typically sell their work on an hourly basis.
Therefore, issues with \textit{Consistent Performance} could be rather problematic.
This highlights that not all roles or companies might be suitable for software engineers with ADHD, as many interviewees stated.

In non-agile environments, M2 considered some of the challenges to be rather problematic.
While the pace might be slower, and thus reduce issues with \textit{Consistent Performance} or \textit{Starting and Finishing}, other challenges might become more problematic, e.g., \textit{Overpromise} and \textit{Dealing with Deadlines} due to the long planning horizon.
Similarly, \textit{Doing Boring Tasks} could cause increased problems, due to a lack of direct reward or feedback.

All managers saw several of the reported strengths as highly relevant in SE.
For instance, \textit{Thinking Ahead} is a skill important to design, architecture, and quality assurance activities (M1), and in general corporate environments (M2).
Also, \textit{Solving Puzzles} is valuable for areas such as security and quality assurance (M1).
\textit{Broad interests/Systems Thinking} is another strength that the managers considered useful, e.g., for backend development (M1) or in startup environments (M2).
In the same direction, M1 and M4 noted that strengths could be especially fostered in people with ADHD, such as providing increasing training on software architecture if they show \textit{Broad Interests/Systems Thinking} (M1), or supporting people with exceptional \textit{Creativity}.
Finally, M3 shared that some of their best workers had ADHD, and their profiles would match with most of the reported strengths.
Clearly, this could be correlation rather than a causation.

In terms of company accommodations, two managers (M3, M4) specifically mentioned that it is feasible.
However, this requires awareness of the different conditions.
Many interviewees were not disclosing their condition for various reasons, e.g., fear of stigma, negative disclosure experiences, or because they viewed ADHD as a ``personal struggle''.
This highlights that companies might have to take the first steps in creating an inclusive environment in which people with ADHD feel safe to discuss their challenges.
In fact, two interviewees shared experiences in which disclosure of their condition led to further employees disclosing or getting diagnosed with ADHD.

\subsection{Recommendations}
In summary, many of the challenges, strengths and strategies we find resonate well with the clinical perspective on ADHD and research on ADHD in the workplace.
However, there are unique perspective introduced by the SE context.
Therefore, we make the following recommendations.

As a first step, companies need to \textbf{invest in awareness regarding neurodiversity} and in \textbf{creating a workplace at which people with ADHD feel safe} disclosing and discussing their condition. Having a company inclusion policy supported by specialized mental health professionals can provide a robust infrastructure to deal with that topic.
Even though not specific to SE, we believe this point is a prerequisite for all further inclusion initiatives.

Second, \textbf{use agile practices that effectively address many of the challenges}.
Specifically, we believe these include \textbf{agile teams} in which group members do \textbf{joint estimation} and agree on \textbf{definitions of done} to avoid over-engineering tasks, \textbf{work on one project/product} only in a \textbf{sustainable pace}, and use various means of direct feedback, such as \textbf{burndown charts and information radiators}. To initiate tasks and fight procrastination, \textbf{pair programming} can be a valuable tool. 

Third, organisations should \textbf{consider alternative career and promotion paths}.
Junior software development positions might be a bad fit for many individuals with ADHD, due to strong deadline-orientation, fine-grained performance evaluations, and repetitive work.
Instead, roles that match and further develop the strength profiles of these individuals have the potential to benefit both the individuals and the organisation.

Finally, organisations should \textbf{keep supporting flexibility in choosing tools, environments, and work modes}. Tools of varied purpose (linters, task management) can provide support and help externalising activities.
Flexibility in, e.g., working from home, using specific software tools of their choice, or work hours benefit most engineers, but seem to be necessary strategies for many people with ADHD. Distractions can compromise focus and productivity. A home office could be a more controlled environment regarding that aspect but companies can also provide more quiet or isolated offices where interruptions or distractions could be avoided.

\section{Conclusion}
In this paper, we contribute to the very limited body of literature about neurodiversity in software engineering by reporting a case study on the experiences of software engineers with ADHD, who are among the possibly largest population of neurodiverse developers.
We interviewed 19 software engineers with ADHD and discussed the results with 4 experienced managers in SE. 

We find challenges, strengths, and strategies that resonate well with existing clinical research on ADHD and research on ADHD in the workplace.
Furthermore, the findings are highly relevant to the SE context. We propose an initial set of recommendations that can improve the inclusion of ADHD developers in software teams. 
While existing agile practices have the potential to address many of the challenges, companies need to take action to better serve this population. This can bring relevant social impact in the lives of neurodiverse developers and can also result in benefits for organisations. By having highly capable individuals being included and lowering their challenges, this leads to higher productivity and better results.

In future work, we plan to study to what extent our findings are specific to software engineers with ADHD as compared to neurotypical engineers or engineers with other neurodivergent subtypes.

\begin{acks}
We would like to thank our anonymous interviewees for being open towards sharing their experiences with us. 
\end{acks}
\bibliographystyle{ACM-Reference-Format}

\end{document}